\documentclass[preprint,amsmath,amssymb,aps]{revtex4-1}
\usepackage{graphicx}
\begin{document}

\title{
Inferring long-range interactions\\
between immune and tumor cells -\\ 
pitfalls and (partial) solutions
}

\author{Claus Metzner}
\email{claus.metzner@gmail.com}
\affiliation{\small 
Biophysics Group, Friedrich-Alexander University Erlangen, Germany
}
\date{\today}

\begin{abstract}
Upcoming immunotherapies for cancer treatment rely on the ability of the immune system to detect and eliminate tumors in the body. A highly simplified version of this process can be studied in a Petri dish: starting with a random distribution of immune and tumor cells, it can be observed in detail how individual immune cells migrate towards nearby tumor cells, establish contact, and attack. Nevertheless, it remains unclear whether the immune cells find their targets by chance, or if they approach them 'on purpose', using remote sensing mechanisms such as chemotaxis. In this work, we present methods to infer the strength and range of long-range cell-cell interactions from time-lapse recorded cell trajectories, using a maximum likelihood method to fit the model parameters. First, we model the interactions as a distance-dependent 'force' that attracts immune cells towards their nearest tumor cell. While this approach correctly recovers the interaction parameters of simulated cells with constant migration properties, it detects spurious interactions in the case of independent cells that spontaneously change their migration behavior over time. We therefore use an alternative approach that models the interactions by distance-dependent probabilities for positive and negative turning angles of the migrating immune cell. We demonstrate that the latter approach finds the correct interaction parameters even with temporally switching cell migration.
\end{abstract}
\maketitle
 
\clearpage
\section*{Introduction}
The human immune system is able to identify and eliminate a huge variety of pathogens \cite{janeway1996immunobiology, kumar2011pathogen}, including bacteria, viruses and fungi, but also transformed body cells, such as tumors \cite{schuster2006cancer, rosenberg2014decade, drake2014breathing}. While this immune response is enormously complex in vivo, basic interactions between immune cells and pathogens can readily be studied in a Patri dish \cite{olofsson2014distinct, zhou2017bystander}: In one of the simplest assays, one starts with a random two-dimensional distribution of, for example, natural killer (NK) and tumor cells. Over a period of a few hours, one can observe in detail how some NKs approach nearby tumor cells, establish steric contact, and attack their targets. And yet it remains difficult to judge whether the NKs find their targets by chance, or if they are chemotactically \cite{eisenbach04} attracted by some molecular traces of the cancer cells.

In this work, we attempt to infer the strength and range of possible long-range cell-cell interactions from measured cell trajectories. For this purpose, we consider an in vitro assay where immune and target cells are randomly mixed together on a plane surface, or within a suitable three-dimensional matrix. We assume that the cells are time-lapse recorded with sufficient temporal resolution. Applying automatic tracking methods to the video recordings provides the individual cell trajectories, which are approximated as chains of straight line segments, connecting the cell positions at discrete time points. Since cell motion in the vertical direction (that is, perpendicular to the Petri dish, where the matrix may be effectively softer due to an open top surface) often differs from the horizontal motion, our method is focusing only on the projected cell migration in the horizontal (x,y) plane.

In our method, we first set up a model that describes both the normal cell migration and the (hypothetical) long-range interaction between immune and tumor cells. We then use Maximum Likelihood Optimization to fit the parameters of the model, in particular the strength and range of the interaction, to the measured cell trajectories. 

On short time scales of a few minutes, cell migration can be modeled as a random walk with a fixed average speed and a fixed degree of directional persistence \cite{codling2008random}. We therefore assume that the step widths $r$ of a cell (defined as the distance between two successive positions in the time-lapse recording) obeys a Rayleigh distribution with scale parameter $w$, and that the turning angle $\theta$ of the cell (defined as the angle between two successive displacement vectors) follows a von Mises distribution with persistence parameter $\kappa$ . 

Possible interactions between an NK and a tumor cell are expected to depend strongly on the distance $R_{NT}$ between the two cells, as the 'cloud' of chemottactractant molecules around the tumor cell decays exponentially \cite{heinrich2017analytical}. For simplicity, we assume here that a given NK cell is - if at all - only affected by the nearest tumor cell at any given time. We furthermore approximate the cells as point-like particles.

\begin{figure}[h!]
\begin{center}
\includegraphics[width=14cm]{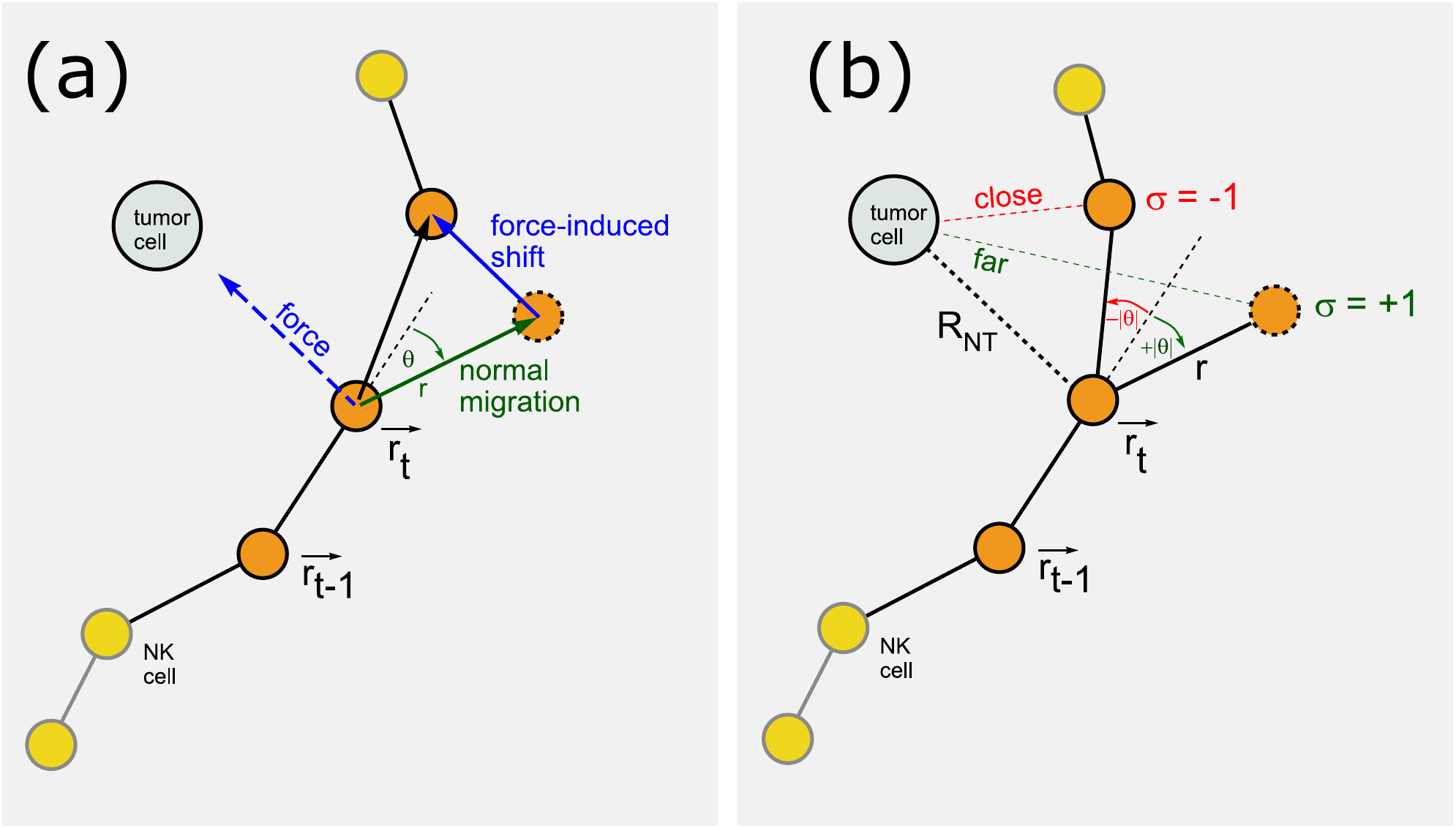}
\caption{
{\bf Illustration of the interaction models}. Shown is a tumor cell (gray circle) and a part of an NK cell trajectory (orange circles with solid borders). The dark orange circles represent a 'triplet' of three successive NK positions. In the force-based model (a), the step of the NK cell between time steps $t$ and $t\!+\!1$ is the sum of the displacements due to regular migration and due to a distance-dependent force towards the tumor cell. In the $\sigma$-based model (b), the NK cell chooses the sign of the turning angle $\theta$ preferentially (that is, with a distance-dependent probability $q$) such that the step leads closer to the tumor cell.
\label{fig_approaches}}
\end{center}
\end{figure}

We start with a physics-inspired approach and describe interactions between the cells by exponentially distance-dependent (presumably attractive) pair forces. Assuming overdamped particle motion, such a force will 'pull' the NK cells (with a velocity proportional to the momentary force vector) towards their nearest tumor cell. However, in contrast to inanimate physical particles, Newton's action-reaction law is invalid here, as tumor cells are of course {\em not} reciprocally attracted towards NK cells. In order to account both for directionally persistent, 'natural' migration of the NK cells, and also for their attraction towards nearby targets, we simply add the displacement vectors of these two contributions in each time step (compare Fig.~\ref{fig_approaches}a). 

This combined migration-interaction model depends on four parameters, namely the most probable step width $w$, the degree of directional persistence $\kappa$, the strength of the interaction force $s$, and its range $d$. Based on the trajectory $\vec{r}_i(t_n)$ of NK cell $i$ and the positions $\vec{r}_j(t_n)$ of all nearby tumor cells $j$, we compute the likelihood $p(step_{n,i}|w,\kappa,s,d)$ of each individual step $n$ of cell $i$, given the parameters of migration-interaction model. The product over all trajectory time steps $n$ and over all NK cells $i$ finally yields the total model likelihood $p(data|w,\kappa,s,d)$ of the observed data set. We can than fit the four parameters using a suitably adapted Maximum Likelihood approach (see Methods section).

A test of this approach with surrogate data shows that the strength $s$ and range $d$ of the cell-cell force, both for attractive and repulsive interactions, can be correctly inferred from simulated trajectories, {\em provided} that the NK cells perform a random walk with constant migration parameters. However, over a typical observation time period of several hours, the migration parameters $w$ and $\kappa$ of the individual cells can fluctuate spontaneously, either abruptly or gradually \cite{Metzner2015}. Since the force-based approach detects interactions based on the NK cell's step widths $r$ and turning angles $\theta$, any change in the statistics of these variables due to a spontaneous fluctuation of $w$ or $\kappa$ may be misinterpreted as evidence for interactions. Indeed, the method finds strong spurious interactions even in simulations where the cells are moving completely independent from each other.    

To avoid this problem, we exploit that the von Mises distribution depends only on $cos(\theta)$, and not on $\sigma\!=\!sgn(\theta)$, the positive or negative sign of the turning angle $\theta$. The quantity $\sigma$ can therefore be considered as an independent dynamical variable that is insensitive to spontaneous fluctuations of $w$ or $\kappa$, but that is related to goal-directed migration: In each step of an NK cell, it has the binary choice between $\sigma\!=\!+1$ and $\sigma\!=\!-1$, and it can select for the option that brings it closer to a target tumor cell (compare Fig.~\ref{fig_approaches}b). Consequently, if the probability $q=prob(\sigma\!\!=\!\!\sigma_{closer})$ of goal-directed choices is larger than $\frac{1}{2}$, the interactions are attractive, while $q<\frac{1}{2}$ corresponds to a repulsive interaction. In our model, we assume that $q$ is a Gaussian function of the distance $R_{NT}$ between the NK and tumor cell, again parameterized by a strength $s$ and a range $d$ (compare Method section). Testing this $\sigma$-based approach with surrogate data shows that now the interaction parameters can be correctly inferred from simulated trajectories, even in cases where the migration parameters of the NK cells are switching randomly.

\section*{Results}
\subsection*{Parameter reconstruction in the force-based approach}

We start with a test of force-based parameter inference. For this purpose, we simulate cell trajectories with the migration-interaction model described above, and try to reconstruct the four model parameters from these trajectories.

The test cases include (a) attractive interactions ($w\!=\!7,\kappa\!=\!3,s\!=\!3,d\!=\!300$), (b) repulsive interactions ($w\!=\!6,\kappa\!=\!4,s\!=\!-2,d\!=\!200$), (c) no interactions ($w\!=\!5,\kappa\!=\!5,s\!=\!0,d\!=\!0$), and (d) no interactions, but migration parameters that switch randomly between two possible values ($w\in\left[5,5/100\right],\kappa\!\in\left[5,5/100\right],s\!=\!0,d\!=\!0)$. Note that in all test cases, the tumor cells are moving independently with migration parameters $w_T\!=4\!$ and $\kappa_T\!=\!0$.

We are not only interested in the inferred parameters $w_{opt},\kappa_{opt}, s_{opt}$ and $d_{opt}$ of maximum likelihood, but also in their statistical spread. For this reason, we present the two-dimensional likelihood distributions $p(w,\kappa\;|\;s\!=\!s_{opt},d\!=\!d_{opt})$ and $p(s,d\;|\;w\!=\!w_{opt},\kappa\!=\!\kappa_{opt})$.

\begin{figure}[h!]
\begin{center}
\includegraphics[width=14cm]{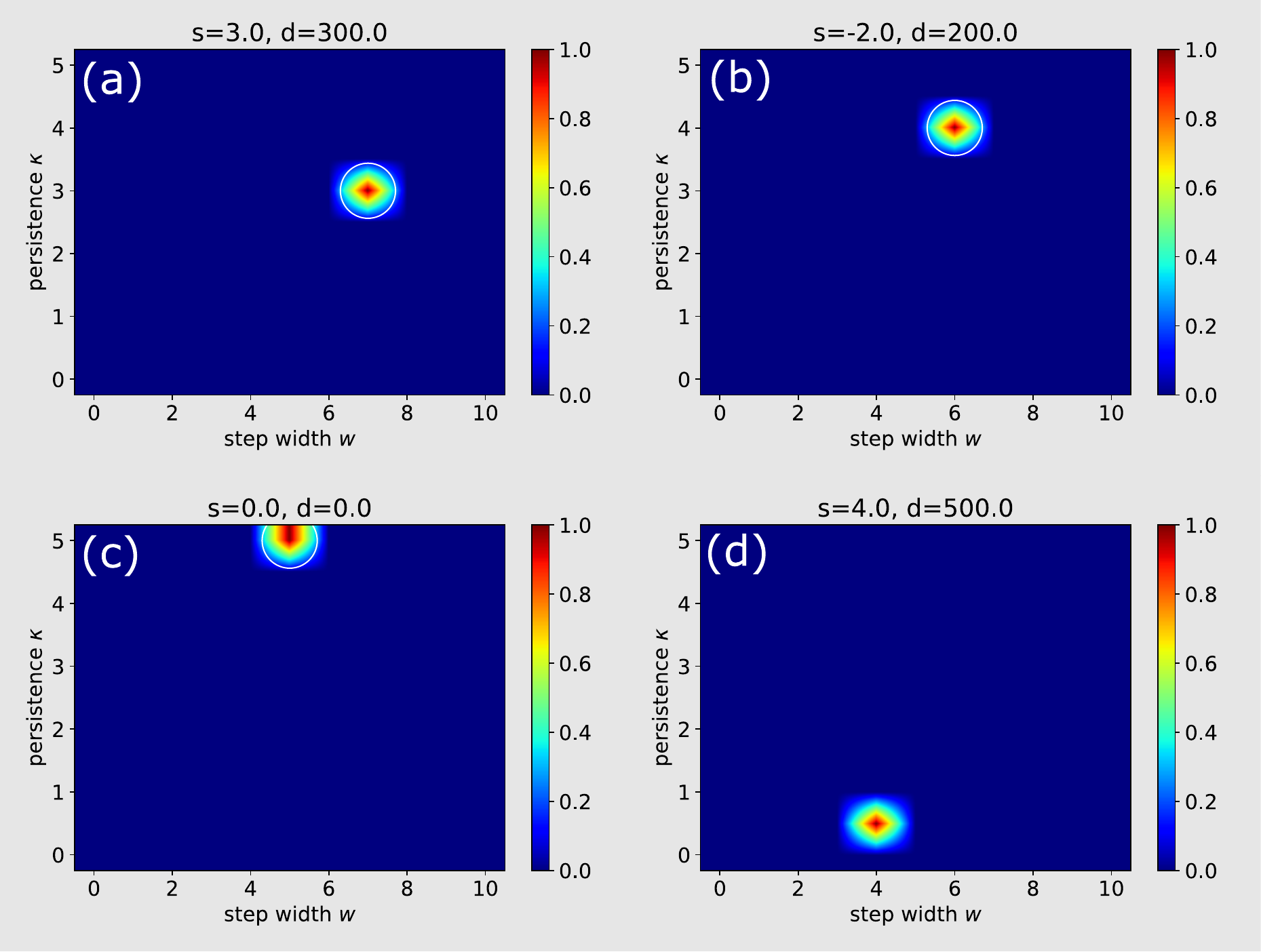}
\caption{
{\bf Force-based model: Reconstruction of migration parameters} (typical step width $w$ and directional persistence $\kappa$) from trajectories that were simulated with the same model. Shown are two-dimensional likelihood distributions $p(w,\kappa | s,d)$ (normalized to a peak value of 1), evaluated for the maximum likelihood interaction parameters (see top of each sub-figure). The true migration parameters (marked by white circles) were (a) $w=7,\kappa=3$, (b) $w=6,\kappa=4$, and (c) $w=5,\kappa=5$. In case (d), migration parameters were randomly switching between two values: $w\in\left\{5,0.05\right\},\kappa\in\left\{5,0.05\right\}$.
\label{fig_frcMig}}
\end{center}
\end{figure}

We find that in all three cases where the simulated migration parameters $w$ and $\kappa$ are constant, their values can be inferred perfectly from the trajectories (Fig.~\ref{fig_frcMig}, a-c). In the case with fluctuating migration parameters (Fig.~\ref{fig_frcMig}, d), the inferred values are somewhere in-between the two alternatives.

\begin{figure}[h!]
\begin{center}
\includegraphics[width=14cm]{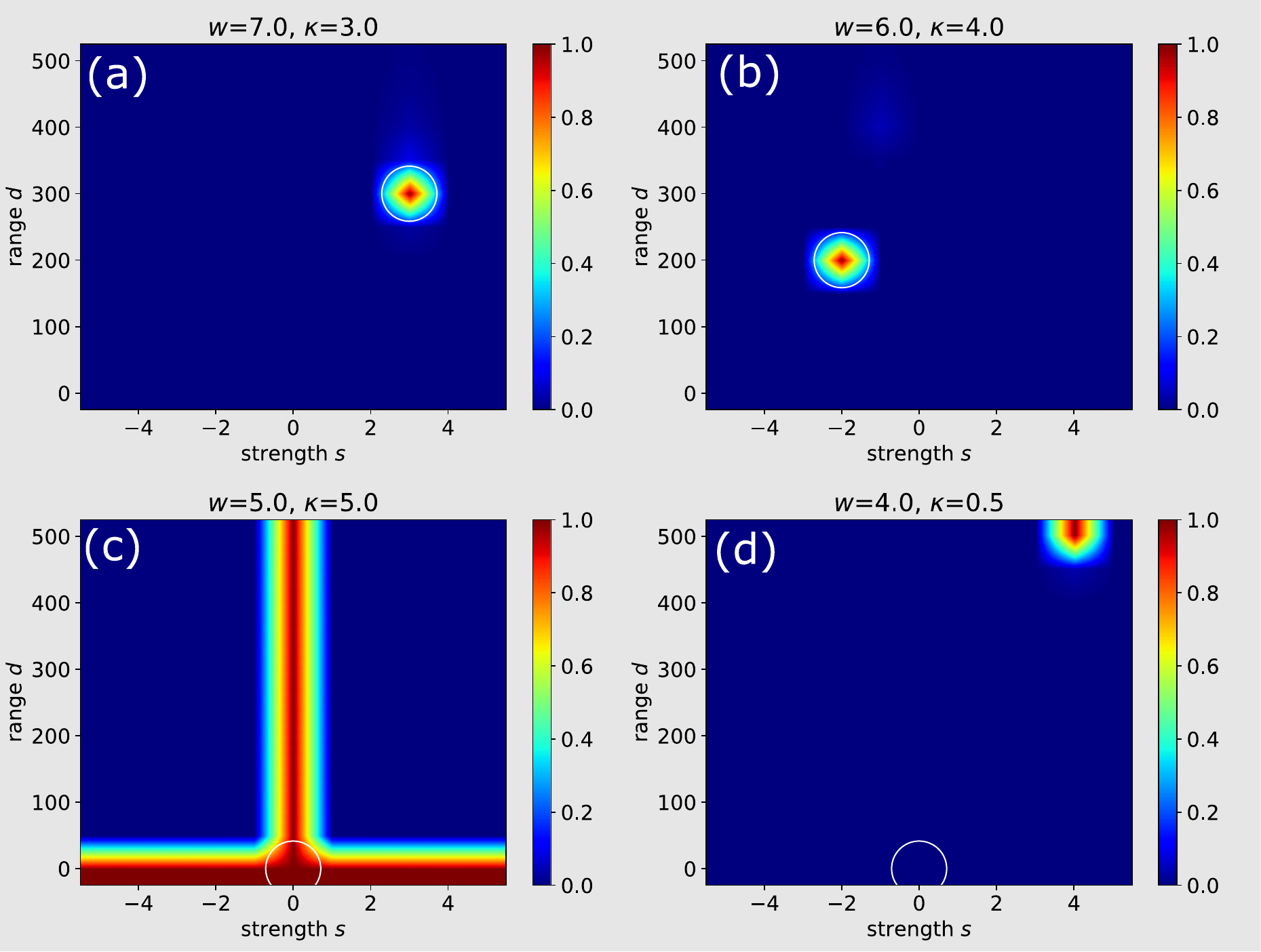}
\caption{
{\bf Force-based model: Reconstruction of interaction parameters} (strength $s$ and range $d$) from trajectories that were simulated with the same model. Shown are two-dimensional likelihood distributions $p(s,d | w,\kappa)$ (normalized to a peak value of 1), evaluated for the maximum likelihood migration parameters (see top of each sub-figure). The true interaction parameters are marked by white circles. In (a), interactions are attractive ($s=3,d=300$). In (b), interactions are repulsive ($s=-2,d=200$). In (c), there are no interactions ($s=0,d=0$) Note that the 'inverted T' shape of the likelihood distribution indicates  the absence of interactions. In (d), there are no interactions ($s=0,d=0$), but migration parameters are randomly switching. The force-based model is finding spurious interactions in this case.
\label{fig_frcInt}}
\end{center}
\end{figure}

The interaction parameters $s$ and $d$ are also perfectly re-constructed from the trajectories, as long as the simulated migration parameters $w$ and $\kappa$ are constant (Fig.~\ref{fig_frcInt}, a-c. Note that the inverted T-shape of the likelihood distribution in case (c) is the characteristic signature of zero interactions). However, in the case with fluctuating migration parameters, the force-based method falsely infers a strongly attractive ($s=4$) and long-range ($d=500$) interaction, although all cells are independent in the simulation (Fig.~\ref{fig_frcInt}, d).

\subsection*{Parameter reconstruction in the $\sigma$-based approach}

Next, we try to reconstruct the four parameters in the $\sigma$-based approach. In order to increase the variety of test cases, the simulated parameters are chosen differently now, including  
(a) repulsive interactions ($w\!=\!2,\kappa\!=\!5,s\!=\!-0.7,d\!=\!300$), 
(b) attractive interactions ($w\!=\!1,\kappa\!=\!3,s\!=\!0.5,d\!=\!200$), 
(c) no interactions ($w\!=\!3,\kappa\!=\!4,s\!=\!0,d\!=\!0$), 
and (d) no interactions, but migration parameters that switch randomly between two possible values ($w\in\left[3,3/100\right],\kappa\!\in\left[4,4/100\right],s\!=\!0,d\!=\!0)$. This time, the tumor cells are moving independently with migration parameters $w_T\!=0.1\!$ and $\kappa_T\!=\!0$.

Note that in the $\sigma$-based approach, the inference of migration and interaction parameters is completely independent. We can therefore plot the non-conditioned two-dimensional likelihood distributions $p(w,\kappa)$ and $p(s,d)$.

\begin{figure}[h!]
\begin{center}
\includegraphics[width=14cm]{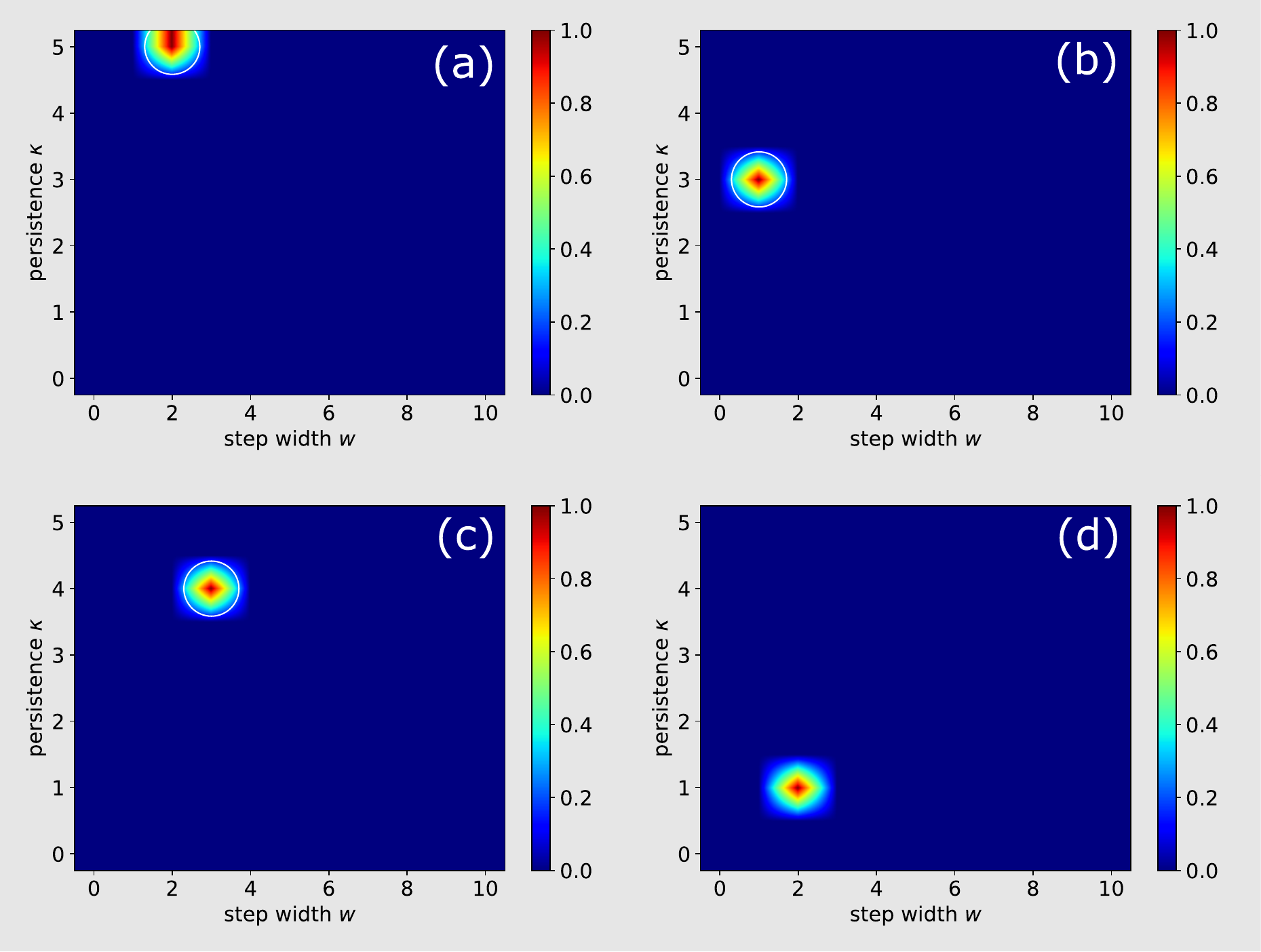}
\caption{
{\bf $\sigma$-based model: Reconstruction of migration parameters} (typical step width $w$ and directional persistence $\kappa$) from trajectories that were simulated with the same model. Shown are two-dimensional likelihood distributions $p(w,\kappa)$, normalized to a peak value of 1. The true migration parameters (marked by white circles) were (a) $w=2,\kappa=5$, (b) $w=1,\kappa=3$, and (c) $w=3,\kappa=4$. In case (d), migration parameters were randomly switching between two values: $w\in\left\{3,0.03\right\},\kappa\in\left\{4,0.04\right\}$.
\label{fig_turMig}}
\end{center}
\end{figure}

Again, we find that in all three cases where the simulated migration parameters $w$ and $\kappa$ are constant, their values can be inferred perfectly from the trajectories (Fig.~\ref{fig_turMig}, a-c). In the case with fluctuating migration parameters (Fig.~\ref{fig_turMig}, d), the inferred values are somewhere in-between the two alternatives.

\begin{figure}[h!]
\begin{center}
\includegraphics[width=14cm]{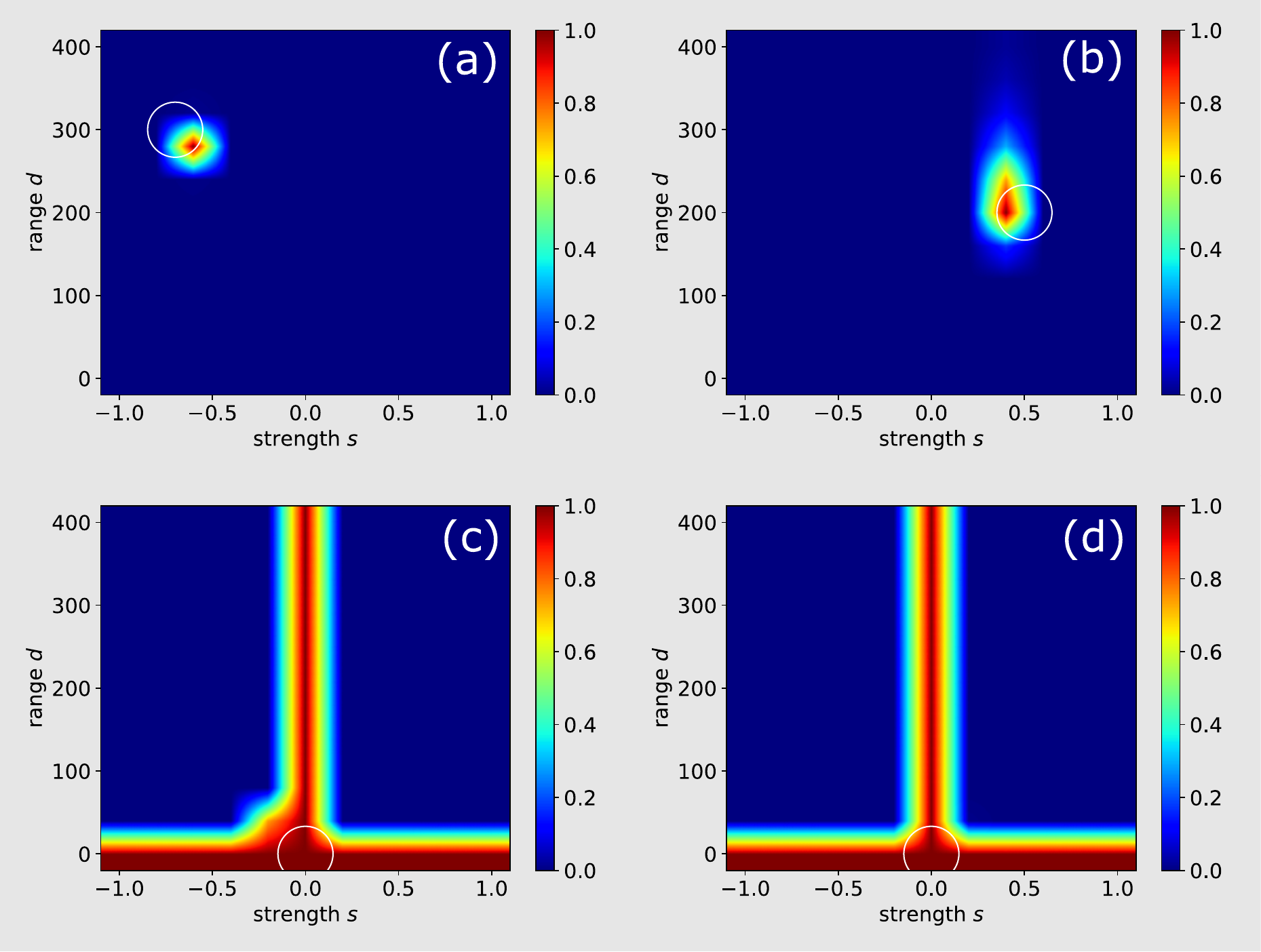}
\caption{
{\bf $\sigma$-based model: Reconstruction of interaction parameters} (strength $s$ and range $d$) from trajectories that were simulated with the same model. Shown are two-dimensional likelihood distributions $p(s,d)$, normalized to a peak value of 1. The true interaction parameters are marked by white circles. In (a), interactions are repulsive ($s=-0.7,d=300$). In (b), interactions are attractive ($s=0.5,d=200$). In (c), there are no interactions ($s=0,d=0$) Note that the 'inverted T' shape of the likelihood distribution indicates the absence of interactions. In (d), there are no interactions ($s=0,d=0$), but migration parameters are randomly switching. The $\sigma$-based model is correctly finding zero interactions in this case.
\label{fig_turInt}}
\end{center}
\end{figure}

The interaction parameters are re-constructed reasonably well in the first three cases (Fig.~\ref{fig_turInt}, a-c). Most importantly, the case (d) with fluctuating migration parameters does not lead to any spurious interactions in the $\sigma$-based approach (Fig.~\ref{fig_turInt}, d). 

\subsection*{Applying the $\sigma$-based approach to chemotaxis simulations}

So far, we have tested our method with simulation data that was generated with exactly the same migration-interaction model as used for parameter inference. However, when applying the method to real world data, it is not clear {a priori} which strategies the immune cells pursue to approach their targets. We therefore need to test our method also with artificial data that are generated with qualitatively different models. 

For this purpose, we use previously published computer simulations of chemotactic 'hunting' behavior \cite{Metzner2019}. These simulations include the following four scenarios:

In 'Blind Search' (BLS), the immune cells do not interact at all with the targets but migrate 'blindly', according to a correlated random walk with fixed parameters for the mean step width (speed) and for the degree of directional persistence. Note that the migration model of these simulations is different from that of the present inference algorithm, since the turning angles are not drawn from a von Mises distribution. We can therefore regard this as an additional test case.

In 'Random Mode Switching' (RMS), the immune cells are still blind with respect to the targets, but occasionally switch between a highly persistent and a non-persistent (diffusive) migration mode. Here, again, the migration model of the simulations is different from that of the present inference algorithm.

In 'Temporal Gradient Sensing' (TGS), the immune cells actually approach the targets by following the temporal gradient of chemo-attractant (For details, see \cite{Metzner2019}). The used model assumes that the immune cells stay in a highly persistent migration mode as long as the concentration of chemo-attractant is increasing with time. When the concentration is decreasing, the immune cells switch to a diffusive mode in order to find a more goal-directed migration direction.

Finally, in 'Spatial Gradient Sensing' (SGS), the immune cells are able to measure the spatial gradient of chemo-attractant and to actively turn into the direction of a nearby target (For details, see \cite{Metzner2019}).

\begin{figure}[h!]
\begin{center}
\includegraphics[width=14cm]{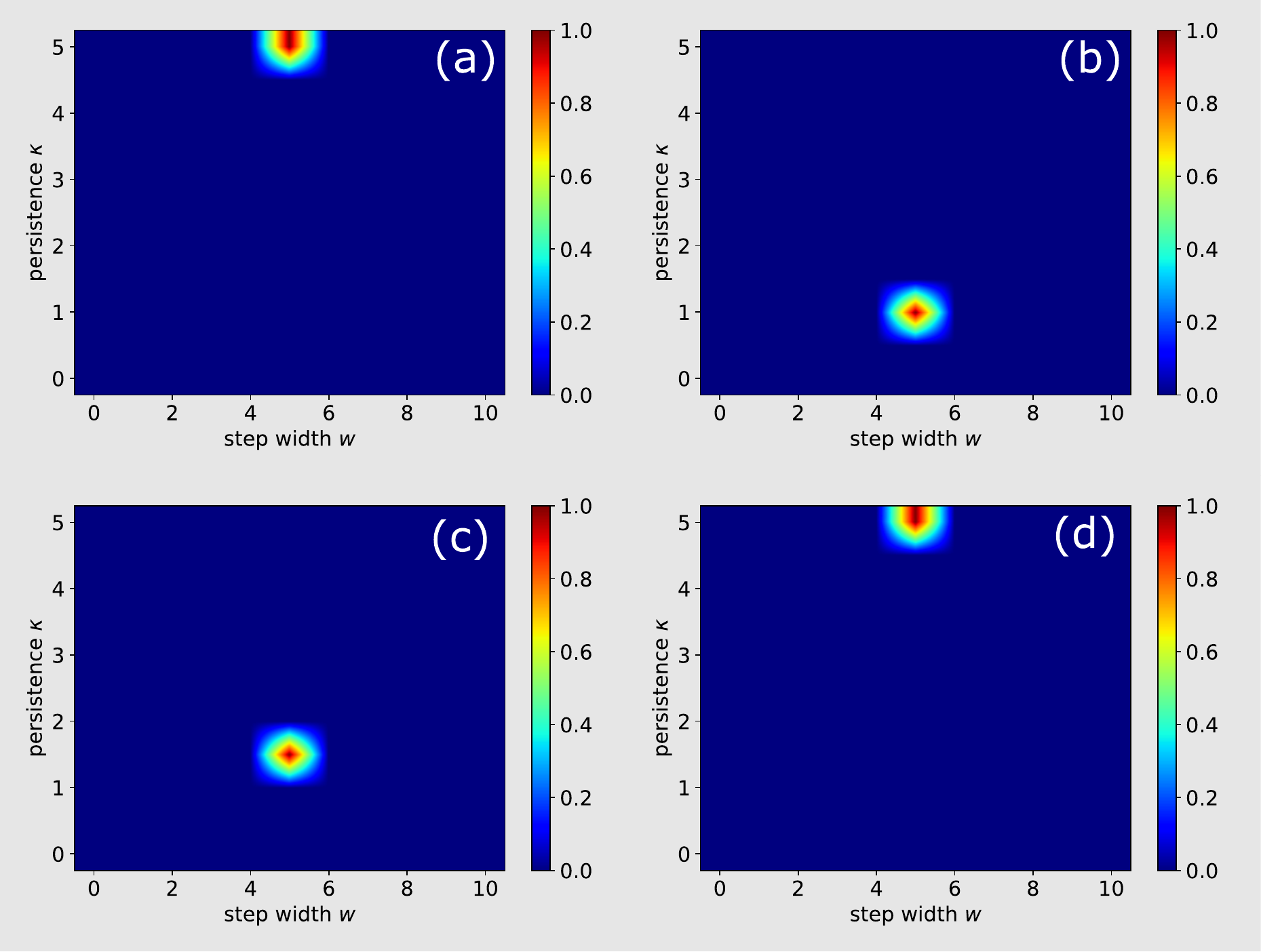}
\caption{
{\bf $\sigma$-based model: Inference of migration parameters} (typical step width $w$ and directional persistence $\kappa$) from trajectories that were simulated with four different models (For details, see Results section). Shown are two-dimensional likelihood distributions $p(w,\kappa)$, normalized to a peak value of 1. Case (a) corresponds to blind search (BLS), case (b) to Random Mode Switching (RMS), case (c) to Temporal Gradient Search (TGS), and case (d) to Spatial Gradient Search (SGS).
\label{fig_turMigChe}}
\end{center}
\end{figure}

When inferring the migration parameters from the above four simulations, we obtain likelihood distributions $p(w,\kappa)$ as shown in Fig.~\ref{fig_turMigChe}. Note that the position of the peak in $p(w,\kappa)$ cannot be compared to the 'true' values, due to the incompatible migration models in simulation and inference. Nevertheless, the inferred values are reasonable: For example, the most probable step width in the BLS simulation was $w\!=\!4.78$, and the persistence was very large (0.9 on a scale from 0 to 1), which corresponds well with Fig.~\ref{fig_turMigChe}(a).  

\begin{figure}[h!]
\begin{center}
\includegraphics[width=14cm]{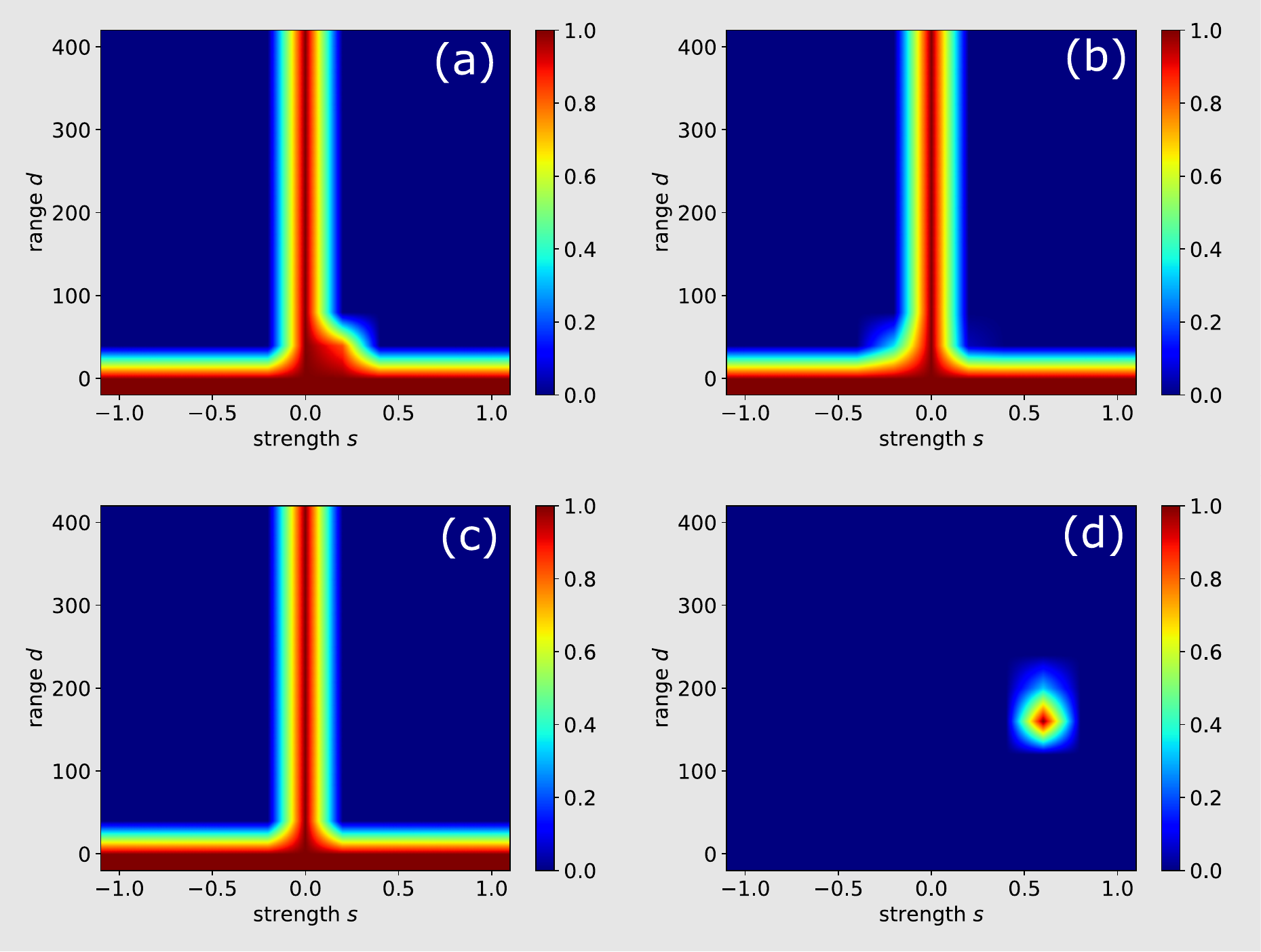}
\caption{
{\bf $\sigma$-based model: Inference of interaction parameters} (strength $s$ and range $d$) from trajectories that were simulated with four different models (For details, see Results section). Shown are two-dimensional likelihood distributions $p(s,d)$, normalized to a peak value of 1. Case (a) corresponds to blind search (BLS), case (b) to Random Mode Switching (RMS), case (c) to Temporal Gradient Search (TGS), and case (d) to Spatial Gradient Search (SGS). The $\sigma$-based model infers correct interaction parameters in three of the four cases (namely BLS, RMS and SGS). However, it fails to recognize the presence of  interactions in the case of the TGS model.
\label{fig_turIntChe}}
\end{center}
\end{figure}

When inferring the interaction parameters from the BLS and RMS simulations (Fig.~\ref{fig_turIntChe}a,b), we find the expected signature of $s\!=\!d\!=\!0$. In the case of SGS (Fig.~\ref{fig_turIntChe}d), we find an attractive interaction, with a strength of about $s\!=\!0.6$ and a range of about $d\!=\!175$. Disappointingly, the $\sigma$-based inference algorithm does not find any interactions in the case of TGS (Fig.~\ref{fig_turIntChe}c). This failure was however to be expected, since in temporal gradient sensing the immune cells are only modulating their degree of directional persistence, depending on the rising or falling of chemoattractant concentration. They do not actually use the sign of turning angles in order to approach their targets.

\section*{Methods}
\subsection*{Quasi-2D and 3D essays}

We assume an experimental assay where immune and cancer cells are mixed together in a collagen gel, or in any other matrix which is suitable for effective cell migration and which enables proper imaging with a microscope. If the matrix layer has a vertical thickness of only a few cell diameters, the system can be considered quasi two-dimensional, and the subsequent analysis can be restricted to the horizontal (x,y) cell positions. In the case of thicker matrices, where two cells can have the same horizontal position but be in different vertical planes, the z-position of the cells has to be measured as well, which is often not possible with very high precision. Moreover, the properties of cell migration often differ between the z- and x-y-directions. For these reasons, out method uses only the horizontal cell coordinates.

\subsection*{Format of input data}

We assume that the cells in a given field of view are time-lapse recorded with sufficient spatial and temporal resolution. Automatic tracking methods can then be used to extract from each video frame the momentary cell configuration, which is stored in a separate file for later convenience. Each configuration file should contain a list of lines in the form  $(x,y,z,i,c)$, with each line corresponding to a specific cell. Here, $x,y,z$ are the coordinates of the cell center, $i$ is an ID number that is unique to each cell and that persists over subsequent video frames, and $c\in\left\{0=immune,1=target\right\}$) is the category of the cell. The number of lines in the configuration files can change from one time point to the next, as cells may leave or enter the microscope's field of view, because of cell division and death, or due to tracking problems. 

\subsection*{Triplet-based analysis}

From the configuration files, we extract the temporal trajectory of each individual cell $i$, defined as the list of 3D positions $\vec{R}^{(i)}_t = (x^{(i)}_t,y^{(i)}_t,z^{(i)}_t)$ for successive time indices $t=0,1,2,\ldots$. In the following, we need only the 2D positions, denoted by $\vec{r}^{(i)}_t = (x^{(i)}_t,y^{(i)}_t)$. 

It is of practical importance that the cells need not to be tracked consecutively over a large number of frames, as our method requires only short 'triplets': sequences of three successive frames in which the positions of the same immune cell $i$ (namely $\vec{r}^{(i)}_{t\!-\!1}$, $\vec{r}^{(i)}_{t}$, and $\vec{r}^{(i)}_{t\!+\!1}$) and of at least one nearby target cell $j$ (namely $\vec{r}^{(j)}_{t}$) are available. If a cell trajectory contains tracking gaps, the specific triplets containing such gaps are excluded from the analysis, but all other triplets are being used.

\subsection*{2D cell migration model}

\vspace{0.2cm}\noindent The sequence of a cell's horizontal positions $\vec{r}^{(i)}_t$ is approximated by a directionally persistent random walk with a certain distribution $p_i(w)$ of step widths $w$, and a distribution $p_i(\theta)$ of turning angles $\theta$. Here, the step width in the move from time $t$ to $t\!+\!1$ is defined as $r=|\vec{r}^{(i)}_{t\!+\!1}-\vec{r}^{(i)}_t|$, and the turning angle $\theta$ is defined as the angle between the two shift vectors $\left[\vec{r}^{(i)}_{t\!+\!1}-\vec{r}^{(i)}_t\right]$ and $\left[\vec{r}^{(i)}_t-\vec{r}^{(i)}_{t\!-\!1}\right]$.

\vspace{0.2cm}\noindent The step width distribution is modeled as a Rayleigh distribution with the most probable step width $w$ (the mode of the distribution):
\begin{equation}
p(r) = \frac{r}{w} \exp\left( -\frac{1}{2}\frac{r^2}{w^2}\right).
\end{equation}

\vspace{0.2cm}\noindent The turning angle distribution is modeled as a von Mises distribution with persistence parameter $\kappa$:
\begin{equation}
p(\theta) = \frac{1}{2\pi I_0(\kappa)} \exp\left( \kappa\cdot\cos(\theta) \right),
\end{equation}
where $I_0$ is the modified Bessel function of order $0$.

\vspace{0.2cm}\noindent For simulating cell migration, new values of $r$ and $\theta$ are drawn from their respective distributions. The 'natural' shift vector of cell $i$ is then given by 
\begin{equation}
\vec{\Delta}_{nat}=\left[\vec{r}^{(i)}_{t\!+\!1}-\vec{r}^{(i)}_{t}\right]_{nat} = r\cdot(cos(\theta),sin(\theta)).
\end{equation}

\subsection*{Force-based interaction model}

Given the position $\vec{r}^{(i)}_t$ of an immune cell $i$ and the position $\vec{r}^{(j)}_t$ of its nearest target cell $j$, the relative vector from $i$ to $j$ is 
\begin{equation}
\vec{u}^{(ij)}_t = \vec{r}^{(j)}_t -\vec{r}^{(i)}_t.
\end{equation}
The magnitude of the force on $i$ is modeled as
\begin{equation}
f^{(i)}_t = s\cdot\exp\left( -|\vec{u}^{(ij)}_t|/d\right),
\end{equation}
where $s$ is the strength parameter and $d$ the range parameter.

\vspace{0.2cm}\noindent The force-induced shift vector of cell $i$ is then given by 
\begin{equation}
\vec{\Delta}_{frc}=\left[\vec{r}^{(i)}_{t\!+\!1}-\vec{r}^{(i)}_{t}\right]_{frc} = f^{(i)}_t\;\left(\vec{u}^{(ij)}_t/\;|\vec{u}^{(ij)}_t|\right).
\end{equation}

\vspace{0.2cm}\noindent Finally, the total shift vector of cell $i$ is computed as the sum of the two contributions: 
\begin{equation}
\vec{\Delta}_{tot}=\left[\vec{r}^{(i)}_{t\!+\!1}-\vec{r}^{(i)}_{t}\right]_{tot} = 
\vec{\Delta}_{nat} + \vec{\Delta}_{frc}.
\end{equation}

\subsection*{$\sigma$-based interaction model}

In the $\sigma$-based interaction model, we assume that the magnitude $|\theta|$ of immune cell $i$'s turning angle and the step width are determined by 'natural' migration, but that $\sigma\!=\!sgn(\theta)\in\left\{-1,+1\right\}$ depends probabilistically on the position of the nearest target cell $j$. In particular, there is a certain probability $q$ that the immune cell will vote for the sign $\sigma$ that leads closer to $j$:
\begin{equation}
q=prob(\sigma\!\!=\!\!\sigma_{closer}).
\end{equation}

\vspace{0.2cm}\noindent We assume that this 'approach probability' $q$ depends on the distance $|\vec{u}^{(ij)}_t|$ to the target as
\begin{equation}
q = \frac{1}{2} + \left( \frac{1}{2}\cdot s\cdot \exp\left( -|\vec{u}^{(ij)}_t|/d\right) \right)
\label{apprProb}
\end{equation}
where $s\in\left[-1,+1\right]$ is a strength parameter and $d$ is a range parameter.

\subsection*{Parameter inference in the force-based model}

In the force-based model, we first compute the total shift $\vec{\Delta}_{tot}$ of the immune cell $i$ between time points $t$ and $t\!+\!1$, using the three successive positions of $i$ that are provided in a triplet.

\vspace{0.2cm}\noindent Next we compute the expected force-induced shift $\vec{\Delta}_{frc}$, given the relative position between the immune and target cell at time $t$, and given some candidate interaction parameters $s$ and $d$.

\vspace{0.2cm}\noindent Next we compute the resulting natural shift $\vec{\Delta}_{nat}=\vec{\Delta}_{tot}-\vec{\Delta}_{frc}$, and decompose it into a step width $r$ and a turning angle $\theta$.

\vspace{0.2cm}\noindent Based on $r$ and $\theta$, and given some candidate migration parameters $w$ and $\kappa$, we compute the logarithm of the likelihood of the step performed by immune cell $i$ in triplet $t$:
\begin{eqnarray}
\left[logLLH\right]^{(i)}_t &=& log\left( p(r,\theta | w,\kappa,s,d)\right)\nonumber\\
&=& log(\frac{r}{w^2})-\frac{r^2}{2w^2}\nonumber\\ 
&+& \kappa\; cos(\theta) - log(2 I_0(\kappa)).
\end{eqnarray}

\vspace{0.2cm}\noindent The total logarithmic likelihood (logLLH) is obtained by summing over all triplets $t$ and all immune cells $i$:
\begin{equation}
log\left(p(data | w,\kappa,s,d)\right) = \sum_{i,t} \left[logLLH\right]^{(i)}_t.
\end{equation}
Note that this logLLH depends on the two migration parameters, but (indirectly) also on the two interaction parameters.

\vspace{0.2cm}\noindent This quantity needs to be maximized over the 4-dimensional space of the model parameters $w$, $\kappa$, $s$, and $d$. For this purpose, we discretize the 4D space with a regular, finite grid (that is, $w_n=w_0+n\cdot \Delta w$, and equivalently for the other parameters). We first set the interaction parameters $s$ and $d$ to arbitrary starting values (on their respective grids) and compute the total logLLH exhaustively for all points on the 2D grid of the migration parameters $w$ and $\kappa$. The resulting optimum values of $w$ and $\kappa$ are fixed and next we exhaustively compute the logLLH on the 2D grid of the interaction parameters $s$ and $d$. Continuing in this way, we alternatingly optimize the migration and interaction parameters. The iteration is stopped when the same set of four optimum parameters is obtained in two successive cycles.

\subsection*{Parameter inference in the $\sigma$-based model}

In the $\sigma$-based model, parameter inference of migration and interaction parameters is completely independent. To extract the {\em migration parameters}, we first compute the total shift $\vec{\Delta}_{tot}$ of the immune cell $i$ between time points $t$ and $t\!+\!1$, using the three successive positions of $i$ that are provided in a triplet. We decompose this shift into a step width $r$ and a turning angle $\theta$. Based on $r$ and $\theta$, and given some candidate migration parameters $w$ and $\kappa$, we compute the logarithm of the likelihood of the step performed by immune cell $i$ in triplet $t$:
\begin{eqnarray}
\left[logLLH\right]^{(i)}_{t,mig} &=& log\left( p(r,\theta | w,\kappa)\right)\nonumber\\
&=& log(\frac{r}{w^2})-\frac{r^2}{2w^2} 
+ \kappa\; cos(\theta) - log(2 I_0(\kappa)).
\end{eqnarray}

\vspace{0.2cm}\noindent The total logarithmic likelihood is again obtained by summing over all triplets $t$ and all immune cells $i$:
\begin{equation}
log\left(p(data | w,\kappa)\right)_{mig} = \sum_{i,t} \left[logLLH\right]^{(i)}_{t,mig}.
\end{equation}
Note that this logLLH depends only on the two migration parameters. It is therefore straight forward to compute it exhaustively on a 2D grid and thus to find the maximum likelihood values of $w$ and $\kappa$. To extract the {\em interaction parameters}, we again compute the actual step width $r$ and turning angle $\theta$ of immune cell $i$ between time points $t$ and $t\!+\!1$. From $\theta$ we extract the actual sign $\sigma=sgn(\theta)$  and, based on the relative vector $\vec{u}^{(ij)}_t$ to the nearest target cell $j$, the sign $\sigma_{closer}$ of the turning angle that would bring $i$ closer to $j$. Given some candidate interaction parameters $s$ and $d$, we furthermore compute the approach probability $q$ according to Eq.~\ref{apprProb}. The logLLH for inference of the interaction parameters is then given by
\begin{equation}
\left[logLLH\right]^{(i)}_{t,int} =
\begin{cases}
    log(q)       & \quad \text{if } \sigma=+\sigma_{closer}\\
    log(1-q)     & \quad \text{if } \sigma=-\sigma_{closer}
  \end{cases}
\end{equation}

\vspace{0.2cm}\noindent As usual, the total logLLH is given by 
\begin{equation}
log\left(p(data | s,d)\right)_{int} = \sum_{i,t} \left[logLLH\right]^{(i)}_{t,int}.
\end{equation}
It is straight forward to compute this quantity exhaustively on a 2D grid and thus to find the maximum likelihood values of $s$ and $d$.

\section*{Discussion}
In this work, we have tested methods to infer the parameters of remote interactions between immune and tumor cells from recorded cell trajectories. In principle, after defining a complete probabilistic model that describes both the normal migration of the cells and the effect of the interactions, it is straightforward to extract the parameters of the model from the data, using Maximum Likelihood estimation. 

In practice, however, this simple task turns out to be more difficult than expected: Our first model, based on distance-dependent forces between the cells, finds spurious interactions in cases where the immune can spontaneously change their migration behavior. Our second model, based on a distance-dependent probability of the immune cells to choose a migration direction that brings it closer to a nearby tumor cells, resolves this problem. This so-called '$\sigma$-based' model correctly recovers all parameters from data that are simulated with the very same model. It also finds reasonable parameters in data from a simulation of chemotaxis that is based on spatial gradient sensing. But even the $\sigma$-based model fails when applied to immune cells that find their targets by modulating their migration parameters as a function of the temporal gradient of chemo-attractant. 

Although we have investigated only two types of interaction models here, our findings point to a more general problem with the inference of interaction parameters: False negative or false positive detections of cell-cell interactions can always occur when a model of cell behavior is assumed for inference that does not even qualitatively reflect the behavior of the actual cells to be analyzed. To reduce the risk of such errors, researchers should at least perform parameter inferences for a sufficiently large set of plausible models.

A better approach may therefore be to consider the detection of remote cell-cell interactions not as a problem of parameter inference, but as a problem of statistical hypothesis testing (for an example, see \cite{Metzner2019b}): Either there exist such interactions, or the null hypothesis is true and the immune cells find their targets by mere chance. Using that approach, the definition of an a-priori interaction model can be totally avoided. 

\bibliographystyle{unsrt}
\bibliography{references}



\section*{Additional information}

\noindent{\bf Author contributions statement:}
CM has devised the study, developed the simulation programs, performed and interpreted the simulations, and wrote the paper.

\noindent{\bf Funding:}
This work was funded by the Grant ME1260/11-1 of the German Research Foundation DFG.

\noindent{\bf Competing interests statement:}
The authors declare no competing interests.

\noindent{\bf Data availability statement:}
All simulation programs and results are available online at {\em 
https://tinyurl.com/infer-interactions}

\noindent{\bf Ethical approval and informed consent:}
Not applicable.

\noindent{\bf Third party rights:}
All material used in the paper are the intellectual property of the authors.

\clearpage
\clearpage


\clearpage


\clearpage


\clearpage


\clearpage


\clearpage


\clearpage


\end{document}